\documentclass[conference]{IEEEtran}
\makeatletter
\def\ps@headings{%
	\def\@oddhead{\mbox{}\scriptsize\rightmark \hfil \thepage}%
	\def\@evenhead{\scriptsize\thepage \hfil \leftmark\mbox{}}%
	\def\@oddfoot{}%
	\def\@evenfoot{}}
\makeatother
\usepackage{listings}
\usepackage{color}
\usepackage{setspace}
\usepackage{float}
\usepackage{subfigure}
\usepackage[fleqn]{amsmath}
\usepackage[utf8x]{inputenc}
\usepackage{multirow}
\usepackage{floatflt}
\usepackage{float}
\usepackage{tabularx}
\usepackage{algpseudocode}
\usepackage{setspace}
\usepackage{mathtools, cuted}
\usepackage{algpseudocode}
\usepackage{setspace}
\usepackage{amssymb}
\usepackage{balance}
\usepackage{graphics}
\usepackage{adjustbox}
\usepackage{mathtools, cuted}
\usepackage{caption}
\usepackage{url}
\usepackage{rotating}

\begin{document}
\title{PhotoSafer: Content-Based and Context-Aware Private Photo Protection for Smartphones}

\author{\IEEEauthorblockN{Ang Li, David Darling, Qinghua Li}\IEEEauthorblockA{Department of Computer Science and Computer Engineering\\ University of Arkansas\\ Email:\{angli,dwdarlin,qinghual\}@uark.edu}}
\maketitle

\begin{abstract}
Nowadays many people store photos in smartphones. Many of the photos contain sensitive, private information, such as a photocopy of driver’s license and credit card. An arising privacy concern is with the unauthorized accesses to such private photos by installed apps. Coarse-grained access control systems such as the Android permission system offer all-or-nothing access to photos stored on smartphones, and users are unaware of the exact behavior of installed apps. Our analysis finds that 82\% of the top 200 free apps in a popular Android app store have complete access to stored photos and network on a user's smartphone, which indicates possible private photo leakage. In addition, our user survey reveals that 87.5\% of the 112 respondents are not aware that certain apps can access their photos without informing users, and all the respondents believe that the stored photos on their smartphones contain different types of private information. Hence, we propose PhotoSafer, a content-based, context-aware private photo protection system for Android phones. PhotoSafer can detect private photos based on photo content with a well-trained deep convolutional neural network, and control access to photos based on system status (e.g., screen locked or not) and app-running status (e.g., app in the background). Evaluations demonstrate that PhotoSafer can accurately identify private photos in real time. The efficacy and efficiency of the implemented prototype system show the potential for practical use.
\end{abstract}


\section{Introduction}
Smartphones have shifted the way people take, store and share photos. There is an increasing number of people who are able to take photos with smartphones anytime, anywhere. Also, almost all social networks allow users to share photos from corresponding smartphone apps (e.g., Instagram). Consequently, more and more people prefer to store photos on their smartphones for convenience, even though some photos are private and sensitive (e.g., driver’s license). It is reported that the average person has 630 photos stored on their phones \cite{howmanyphotos}. However, many installed apps on smartphones have access to stored photos and the network, which may cause leakage of private photos to remote parties. This raises a privacy concern that users’ private photos might be accessed by apps without their awareness.

The Android platform offers users two approaches for controlling access permissions. At the early stage of Android, users are asked to grant permissions when they install an app. Specifically, an app will disclose the full list of resources that it wants to access at installation. Either all requested permissions are granted, or the entire installation is aborted. Prior research has shown that most users do not care about or understand these disclosures at installation \cite{felt2012android}. With the evolution of the Android platform, a new permission scheme has replaced the install-time disclosures for enhancing smoothness of installation process. In particular, users need to grant permissions only when an app requests a sensitive resource for the first use. Their decisions to these permission requests will be applied to all future requests by that app for the same permission. However, this scheme only considers a user’s preference for permission requests when an app is used for the first time. An app once granted access to a photo at the first access will be able to access all photos all the time. It does not account for the fact that the user's preference for subsequent permission requests might change under different contextual circumstances. For instance, a user is willing to upload a photo that was taken in a private gathering through a social-network app; however, the same user might feel uncomfortable for the same app running in the background to access such private photos without his awareness.

To protect private photos, some apps have been developed \cite{gallerylock,photovault,privatevault}, which apply authentication techniques (e.g., password and fingerprint) to control access to those photos. However, they either significantly affect the usability and user experience or cannot really secure private photos. Specifically, users are usually required to manually identify and import private photos from the native photo gallery app on Android to such third-party apps. It is very challenging and boring for users to manually select private photos from hundreds or even thousands photos on their mobile phones. Moreover, some of these apps only copy private photos to a specific protected folder but still keep them in the native photo gallery app, which requires users to remove those private photos from the native photo gallery app. If a user forgets to do so, no protection can be provided. Even worse, some apps merely move user-specified private photos to a hidden folder, which can be easily detected and accessed without any challenge by using existing file management apps \cite{filemanager,esfile,filemangement}. In addition, when a user wants to share private photos with other people through social network apps such as Instagram, since these social network apps usually only allow users to choose to-be-shared photos from the native gallery app or the file management system, it is inconvenient for users if private photos are kept in separate app-specified folders. Hence, existing solutions cannot really secure private photos while offering friendly user experience.

Some work has been done for refining Android permission systems. Nauman et al. \cite{nauman2010apex} and Jeon et al. \cite{jeon2012dr} designed fine-grained permissions for Android, but do not specifically protect stored private photos. CHIPS \cite{tan2014short} is a face-recognition-based access control system for stored photos on Android phones, but can only protect photos that contain pre-specified faces, which cannot be applied to other types of private photos (e.g., credit card).

To this end, we design a novel content-based, context-aware private photo protection system named \textit{PhotoSafer} for Android phones, which provides real-time access control over private photos based on photo contents and the contextual status of accesses, and discloses the specific sensitive content that private photos contain to users before the photos can be accessed. Our contributions are as follows:
\begin{itemize}
\item We analyze the top 200 free apps on Apkpure, which is a very popular third-party Android app store, for evaluating the potential privacy risks that current apps pose to private photos.
\item We conduct an online survey with 112 respondents to investigate smartphone users' privacy concerns over private photos, including common types of private photos, awareness of photo-accessing operations by apps, etc.
\item We design a novel content-based, context-aware private photo protection system PhotoSafer, which can automatically identify private photos and perform real-time access control over private photos based on the contextual status of mobile phone and whether the requesting apps are running in the foreground.
\item We implement a prototype system on the Nexus 5 phone, and evaluate the system's performance.
\end{itemize}

The rest of the paper is organized as follows. Section \ref{sec:motivation} presents how this work is motivated, including a permission analysis of 200 popular apps and an online survey. Section \ref{sec:systemdesign} introduces the design and workflow of PhotoSafer. Section \ref{sec:implementation} describes the prototype implementation. Section \ref{sec:evaluation} shows evaluation results. Section \ref{sec:discussion} discusses the limitations of this work. Section \ref{sec: relatedwork} reviews related work. Section \ref{conclusion} concludes the paper and discusses future work.

\section{Motivation}\label{sec:motivation}
To better understand the privacy issues with photos stored on mobile phones, we firstly analyze the requested permissions of 200 apps to demonstrate the potential risk of unauthorized access to private photos, and then investigate users' concerns about private photos in the real world through an online survey.

\subsection{Permission Analysis}
Let us first analyze what permissions are required to access stored photos on the Android platform. For an Android device, photos are stored in the external storage directory that can be either a physical removable memory card or a logical partition in the device's memory. Hence, to access stored photos, an app has to be granted the permission  
\texttt{READ\_EXTERNAL\_STORAGE}, which is the only required permission. However, the external storage directory is not the repository only for photos, but also for other files such as songs. As a result, the correlation between the permission  \path{READ_EXTERNAL_STORAGE} and photo access control is not intuitive to average users. In addition, due to the aforementioned limitations of the Android permission system, users are allowed to choose whether an app can access to all stored photos, but cannot do selective control over any individual photo.

Next, we analyze apps' requested permissions to examine the potential risk of unauthorized access to stored photos. Due to the download restrictions of Google Play, we analyze the top 200 free apps (e.g., Facebook, Twitter, etc.) from Apkpure \cite{apkpure}, which is a popular third-party Android app store. We particularly identify apps that request both \texttt{READ\_EXTERNAL\_STORAGE} and \texttt{INTERNET} permissions, since the combination of these two permissions allow potential leakage of private photos to another party. The analysis tool \textit{Androguard} \cite{desnos2011androguard} is used to extract the requested permissions of each analyzed app. It is found that 164 out of the 200 apps request both \path{READ_EXTERNAL_STORAGE} and \texttt{INTERNET} permissions. That means 82\% of the top 200 free apps on Apkpure have complete access to stored photos on a user's phone, and could even leak these photos through the Internet. Thus, there is a necessity for finer-grained access control on private photos.

\subsection{Online Survey}
PhotoSafer's design is also motivated by an online survey which is designed to investigate smartphone users' concerns about unauthorized access to private photos. The survey was conducted with user consent under an IRB approval from the University of Arkansas. The survey is available online \cite{survey}, and the results here show statistics of all the 112 responses collected by December 22, 2017.  The mobile phone platform usage of respondents is described in Table \ref{tb:phone}, and the age distribution of survey respondents is shown in Table \ref{tb:age}.

\begin{table}[h]
\centering
\caption{Mobile Phone Platform Usage of Survey Respondents}
	\begin{tabular}{c|c}
		\hline
		\textbf{Mobile Phone Platform} & \textbf{Proportion of Respondents}\\
		\hline
		Android & $77.7\%$ \\
		iPhone & $21.4\%$ \\
		Windows Phone & $0.9\%$ \\
		\hline
	\end{tabular}\label{tb:phone}
\end{table}

\begin{table}[h]
	\centering
	\caption{Age Group Distribution of Survey Respondents}
	\begin{tabular}{c|c}
		\hline
		\textbf{Age Group} & \textbf{Proportion of Respondents} \\
		\hline
		Less than 20 years & $5.4\%$ \\
		20-30 years & $84.8\%$ \\
		30-40 years & $9.8\%$ \\
		\hline
	\end{tabular}\label{tb:age}
\end{table}

Participants were asked  whether they store private photos (driver's license, passport, etc.) on their mobile phones. An overwhelming majority (88.6\%) deemed that some private photos are stored on their mobile phones. To explore which specific type of photos are considered as private by respondents, this survey provides different options for participants. As shown in Table \ref{tb:content}, almost every participant considered photos that contain  \textit{Photo ID}, \textit{Legal Documents}  and \textit{Family} as private, and over a half (57.9\%) agreed that nude photos are also sensitive. In a consequence, the above four types of photo contents are used as references to identify different categories of private photos for this work. Even though those types do not cover all cases, they can represent a significant portion of private photos in the real world.

\begin{table}[h]
	\centering
	\caption{Photo Types and the Proportion of Respondents That Consider Them as Private}
	\resizebox{0.49\textwidth}{!}{
	\begin{tabular}{c|c}
		\hline
		\textbf{Photo Type} & \textbf{Proportion of Respondents} \\
		\hline
		Photo ID (e.g., driver's license, passport) & 97.4\% \\
		Legal Documents (e.g. SSN) & 97.4\% \\
		Family (e.g., family party) & 76.5\%\\
		Nudes & 57.9\%\\
		\hline
	\end{tabular}}\label{tb:content}
\end{table}

Furthermore, as shown in Table \ref{tb:apps}, 67.5\% of the participants agreed that there are more than 10 installed apps on their mobile phones that are granted access to photos. Also, for each of the participants, there is at least one installed app that has access to the photos stored on her/his mobile phone. However, the results show that most of participants (87.5\%) do not clearly know whether any installed app can access photos in the background or not without their awareness. As a result, it is an urgent necessity to design a system to protect private photos from being accessed without users' awareness.

\begin{table}[h]
	\centering
	\caption{Number of Installed Apps That Can Access Photos}
	\begin{tabular}{c|c}
		\hline
		\textbf{Number of Installed Apps} & \textbf{Proportion of Respondents} \\
		\hline
		10+ & 67.5\% \\
		6-10 & 21.9\% \\
		1-5 & 10.4\% \\
		\hline
	\end{tabular}\label{tb:apps}
\end{table}

\section{System Design}\label{sec:systemdesign}
This section describes the design of PhotoSafer.

\subsection{System Overview}
Our goal for designing PhotoSafer is to protect private photos from unauthorized access by mobile apps, without changing the way apps access photos and how users store photos on their mobile phones. In addition, the system should not affect the usability of apps and user experience; i.e., the access control enforcement decision must be made within a reasonable amount of time.

Our basic idea is that when an app requests to access a particular photo, users should be aware of it and decide whether the app can access that photo. The naive approach is to prompt users to check the photo and make a decision every time. However, this will definitely degrade the usability of the system and apps. To address this problem, PhotoSafer is designed to be able to automatically check whether the content of photo is private, and determine whether the user is aware of the app's access request based on the contextual status of the phone and the app. Specifically, when the phone is locked, the user is not operating the phone for photo access and thus most likely does not know that an app is accessing the photo. Even when the phone is unlocked, the app which requests access to the photo might be running in the background. In this case, the user probably also does not know that the app is accessing the photo. Generally, a user is aware of the photo access if the app is running in the foreground, since usually the access request is triggered by the user in this scenario. In this case, the system can automatically check the photo content through a trained classifier and inform the user whether the photo contains private information and what private information it is. The system also allows the user to determine whether the access request should be permitted. To minimize the time needed to identify private photo content during user operation, PhotoSafer caches the identified contents of photos in a database in advance. In this way, PhotoSafer can achieve real-time response to photo access, such that the requesting app's usability and user experience will not be affected. To make the system work, PhotoSafer needs to be integrated into the Android kernel as a system service, since it needs privileges to interpose photo access. When PhotoSafer is initialized, it will feed all stored photos into a trained classifier to identify photo contents (e.g., photo ID), and then the result will be stored in a database. Whenever a new photo is added, it will be fed into the classifier and the classification result will be updated into the database.

\begin{figure}[h]
	\centering
	\includegraphics[scale=0.35]{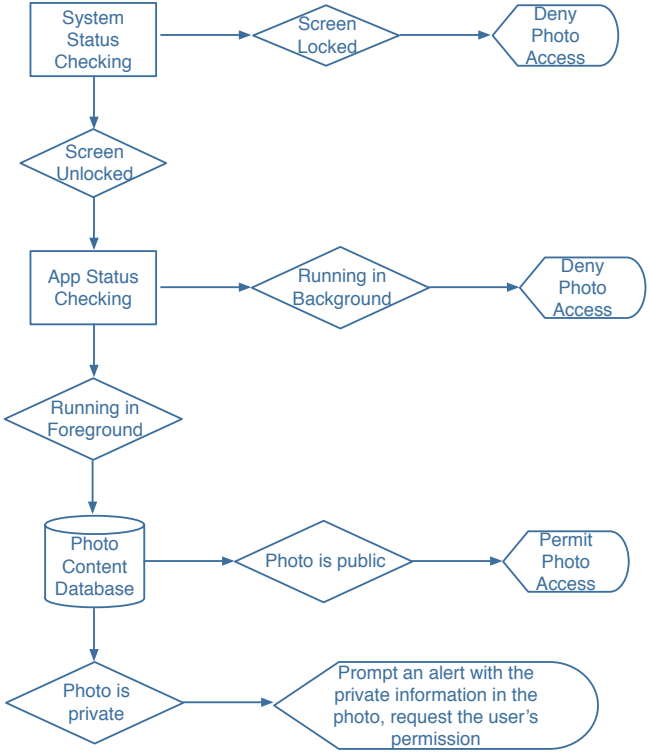}
	\caption{The workflow of PhotoSafer}\label{fig:workflow}
\end{figure}

The workflow of PhotoSafer is shown in Figure \ref{fig:workflow}. When an app requests to access a specific photo, the photo access will be interposed and system status will be checked. If the phone is locked, then the access request will be automatically denied. However, if the phone is unlocked, the system will continue checking the app's status. If the app that requests photo access is running in the background, then the access request will also be automatically denied. On the contrary, if the app is running in the foreground, the photo content can be immediately obtained by querying the database, where the content type of each photo is stored. Finally, if the photo is classified as public (i.e., the photo does not contain private content), the access permission will be automatically granted. Otherwise, an alert will be prompted by informing the user of what private information is contained in the photo and requesting the user to determine permission. If the user trusts the  app and grants permission to it, then the photo access will be continued; otherwise it will be denied.

\subsection{Architecture}\label{subsec:architecture}
\begin{figure}[h]
		\centering
		\includegraphics[scale=0.5]{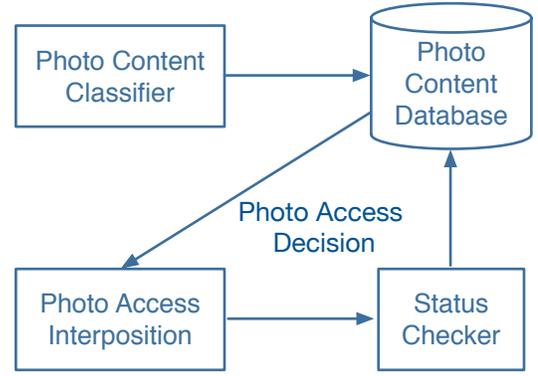}
		\caption{The architecture of PhotoSafer}\label{fig:architecture}
	\end{figure}

As Figure \ref{fig:architecture} shows, the system consists of four major modules: \textit{photo access interposition}, \textit{status checker}, \textit{photo content classifier}, and \textit{photo content database}. We can divide the overall workflow of PhotoSafer into three steps. First, when the system is initialized on a phone, the pre-trained photo content classifier performs classifications on all stored photos, and the results will be stored in the photo content database. In the database, each record consists of a tuple $(photo\_id, content\_type)$, where $photo \_id$ is the unique identifier in each photo's universal resource identifier (e.g., \url{content://com.android.providers.media.documents/document/image%photo_id}) in Android and $content\_type$ represents whether a specific  photo is not private or contains which specific type of private content, such as (`$10001$', `public') and (`$10002$', `photo ID'). Then, when an app requests to access a specific photo, the photo access interposition module will interrupt the app's operation and trigger the status checker module to check the  system status and  the app's  current running status. If the phone is either locked or the app is running in background, the access request will be directly denied. Otherwise, a query with the photo's $photo \_id$ will be sent to the photo content database. Finally, if the returned result from photo content database shows the photo is public, the access request will be automatically permitted, and photo access interposition will resume the app's operation. However, if the photo contains some private information,  an alert will be prompted by describing what private information is contained in the photo and requesting the user to make the decision of whether the photo access interposition should resume the app's operation or not.	
	
The design of PhotoSafer is based on several technologies available in off-the-shelf mobile phones. Photo access interposition can be done by modifying the Android kernel. The status checker can be implemented by Android APIs \texttt{KeyguardManager} and \texttt{ActivityManager}. The photo content database can be implemented by using SQLite. Next, we will describe how to identify private photos.

\subsection{Photo Content Classifier}\label{subsec:design_classifier}
Given an input photo in pre-defined dimensions, this module aims to detect whether that photo contains some specific private information. This is done by training a deep convolutional neural network (DCNN) to detect the private content of photos.

Formally, let $\mathbb{P}$ be the set of input photos. For a given photo $x \in \mathbb{P}$, let $y \in \{1, 2, 3, 4, 5\}$ encode the categorical labels \{`public', `photo id', `legal document', `family', `nude'\} of the photo. Let $\mathbb{H}$ be the hypothesis space of possible decision functions, and $f(\theta^{T}x)$ be the decision function, where $\theta=\{\theta_1,\theta_2,\ldots,\theta_N\}$ is the network weights. Hence, the loss function  can be defined as $L(f(\theta^{T}x),y)$. Let $\mathbb{E}(L)$ be the expected loss over the range of inputs $\mathbb{P}$. In this work, we use cross-entropy to estimate the loss, and hence the optimization task is to minimize the expected cross-entropy loss.

\begin{equation}\label{equ1}
 f=\underset{f\in \mathbb{H}}{\arg\min} \hspace{0.1cm}\mathbb{E}(L)
\end{equation}

For each input $x$, the corresponding classification result is $f(x)$, and hence the according accuracy $acc(x)$ can be defined as:

\begin{equation}
acc(x,y)=
\begin{cases}
+1 & y=f(x)\\
0 & \text{otherwise}
\end{cases}
\end{equation}

However, the main challenge for training a DCNN to identify private photos is acquiring a sufficient number of private photos to train on. Generally, a DCNN requires a relatively large set of training data to perform well. To address this challenge, we adopt the transfer learning \cite{pan2010survey} approach to train our DCNN model. Specifically, we pretrain a DCNN model on a large dataset ImageNet \cite{krizhevsky2012imagenet}, which contains 1.2 million images with 1000 categories. Then, we tune the parameters of the output layer in the pretrained model on a smaller number of private photos that we have collected.

\section{Implementation}\label{sec:implementation}
Due to the time limitation, we implemented PhotoSafer as a standalone app on Android phones instead of integrating it into the Android kernel. We plan to implement its integration with Android kernel in future work. Generally, the app works like the native photo gallery app that comes with the Android system. The prototype app was specifically designed so that it will access some private photos under different system status and app-running status. The photo content classifier was implemented using Python 2.7 and Tensorflow \cite{abadi2016tensorflow}, which is an open-source deep learning framework. The other modules were implemented by available technologies, as mentioned in Section \ref{subsec:architecture}.

\subsection{Photo Content Classifier}
This module aims to identify whether a given photo is public or contains some specific private information. As described in Section \ref{subsec:design_classifier}, we use transfer learning to train a DCNN model. In particular, we build the classifier using the Python APIs of Tensorflow and adopting MobileNets \cite{howard2017mobilenets}. The MobileNets are a class of DCNNs that are specifically designed for efficiently running on mobile devices. The significant difference between the MobileNets architecture and a traditional DCNN’s is that instead of a single 3x3 convolution layer followed by batch norm and rectified linear unit (ReLU), MobileNets split the convolution into a 3$\times$3 depthwise convolution layer and a 1$\times$1 pointwise convolution layer. It has been demonstrated that the computing operations and model size will be significantly reduced in this way. MobileNets are usually not as accurate as traditional DCNNs, but it provides a trade-off between accuracy and resource usage. Specifically, MobileNets offer two parameters to tune the resource and accuracy trade-off: \textit{width multiplier} and  \textit{resolution multiplier}. The value of width multiplier should be set between 0 and 1, while the resolution multiplier might be various. The width multiplier allows us to adjust the thickness of the DCNN, and  the resolution multiplier changes the input dimensions of images, which can reduce the internal representation complexity at every layer. Table \ref{tb:width} shows that, given a fixed resolution multiplier, when the width multiplier increases the number of computing operations and parameters also dramatically increases. However, when the width multiplier is fixed, the larger the input dimension, the more the required computing operations.

\begin{table*}[h]
\footnotesize
	\centering
	\caption{MobileNets with Different Width Multipliers}
	\resizebox{\textwidth}{!}{
	\begin{tabular}{l|c|c|c}
		\hline
		\textbf{Width Multiplier} & \textbf{ImageNet Accuracy} & \textbf{Million Operations of Mult-Add} & \textbf{Million Parameters} \\
		\hline
		MobileNet\_1.0\_224 & 70.6\% & 569 & 4.2 \\
		MobileNet\_0.75\_224 & 68.4\% & 325 & 2.6 \\
		MobileNet\_0.5\_224 & 63.7\% & 149 & 1.3 \\
		MobileNet\_0.25\_224 & 50.6\% & 41 & 0.5 \\
		\hline
	\end{tabular}}\label{tb:width}
\end{table*}

\begin{table*}[h]
\footnotesize
	\centering
	\caption{MobileNets with Different Resolution Multipliers}
	\resizebox{\textwidth}{!}{
		\begin{tabular}{l|c|c|c}
			\hline
			\textbf{Resolution Multiplier} & \textbf{ImageNet Accuracy} & \textbf{Million Operations of Mult-Add} & \textbf{Million Parameters} \\
			\hline
			MobileNet\_1.0\_224 & 70.6\% & 569 & 4.2 \\
			MobileNet\_1.0\_192 & 69.1\% & 418 & 4.2 \\
			MobileNet\_1.0\_160 & 67.2\% & 290 & 4.2 \\
			MobileNet\_1.0\_224 & 64.4\% & 186 & 4.2 \\
			\hline
	\end{tabular}}\label{tb:resoulution}
\end{table*}

In this work, we fix the input dimension as $224\times224$, but change the width multiplier for comparisons in Section  \ref{sub:accuracy}. Firstly, we train the MobileNets on ImageNet with fine-tuning parameters. After that, we fine-tune the output layer of pretrained model with our collected dataset of private photos, but keep the parameters of other layers unchanged. The details of the dataset are described in Section \ref{subsec:dataset}.

\section{Evaluation}\label{sec:evaluation}
In this section, we systematically evaluate the performance of PhotoSafer through a number of experiments. To better illustrate the benefits provided by our proposed system, we also make comparisons against existing approaches. Particularly, we conduct the following experiments. Firstly, we conduct extensive experiments to measure the private photo identification accuracy. Secondly, we  test the time taken by the system to obtain photo content classification results from the database. The evaluations for classification accuracy  are done on Ubuntu 17.04 64-bit machine with 32G RAM and one NVIDIA TITAN Xp GPU. The other experiments are conducted on Nexus 5 phones.

\subsection{Dataset}\label{subsec:dataset}
\begin{figure}[]
\centering
\subfigure[Driver's License]{
\centering
\includegraphics[scale=0.04]{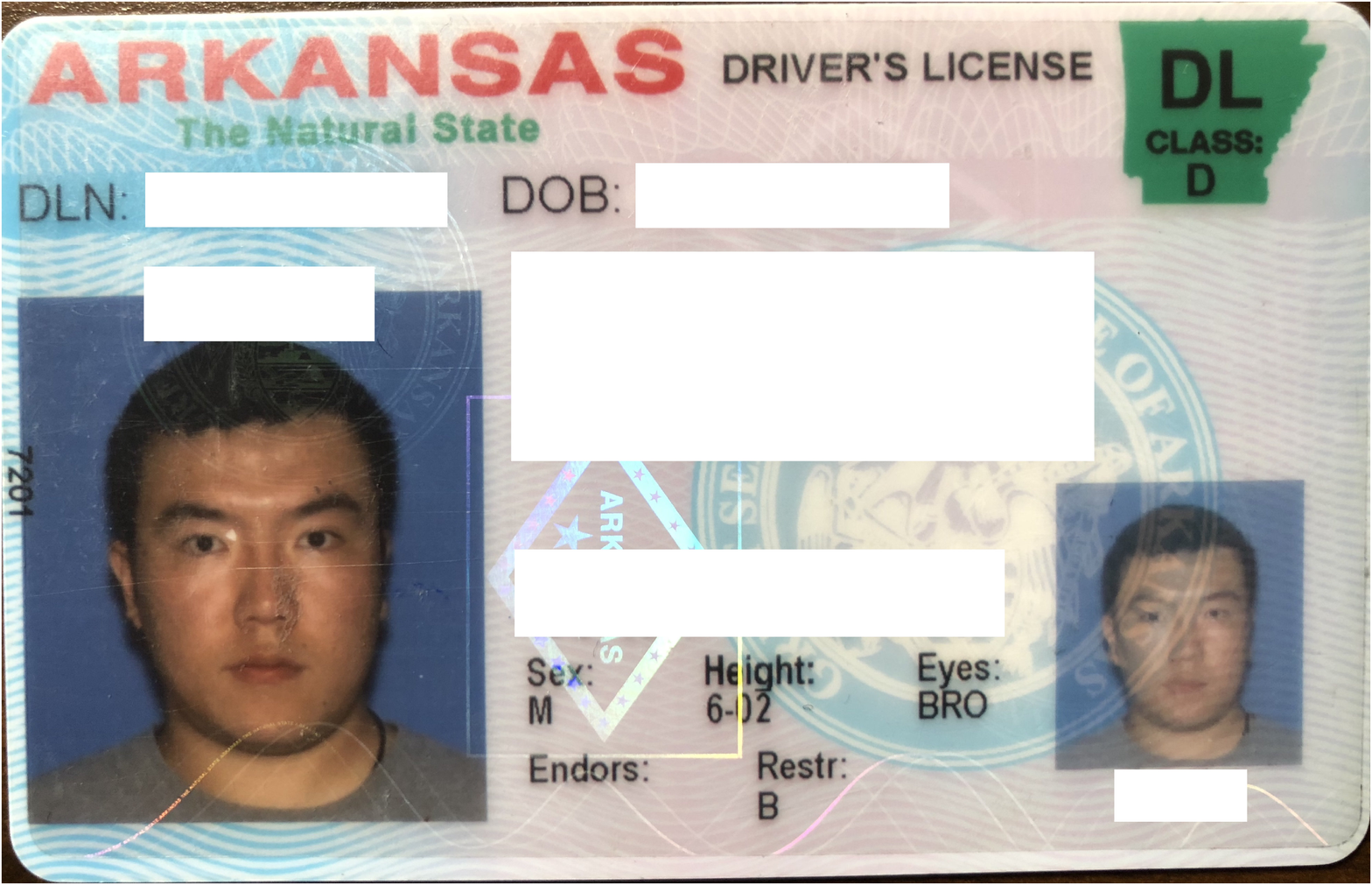}
}
\subfigure[Photo ID]{
\centering
\includegraphics[scale=0.13]{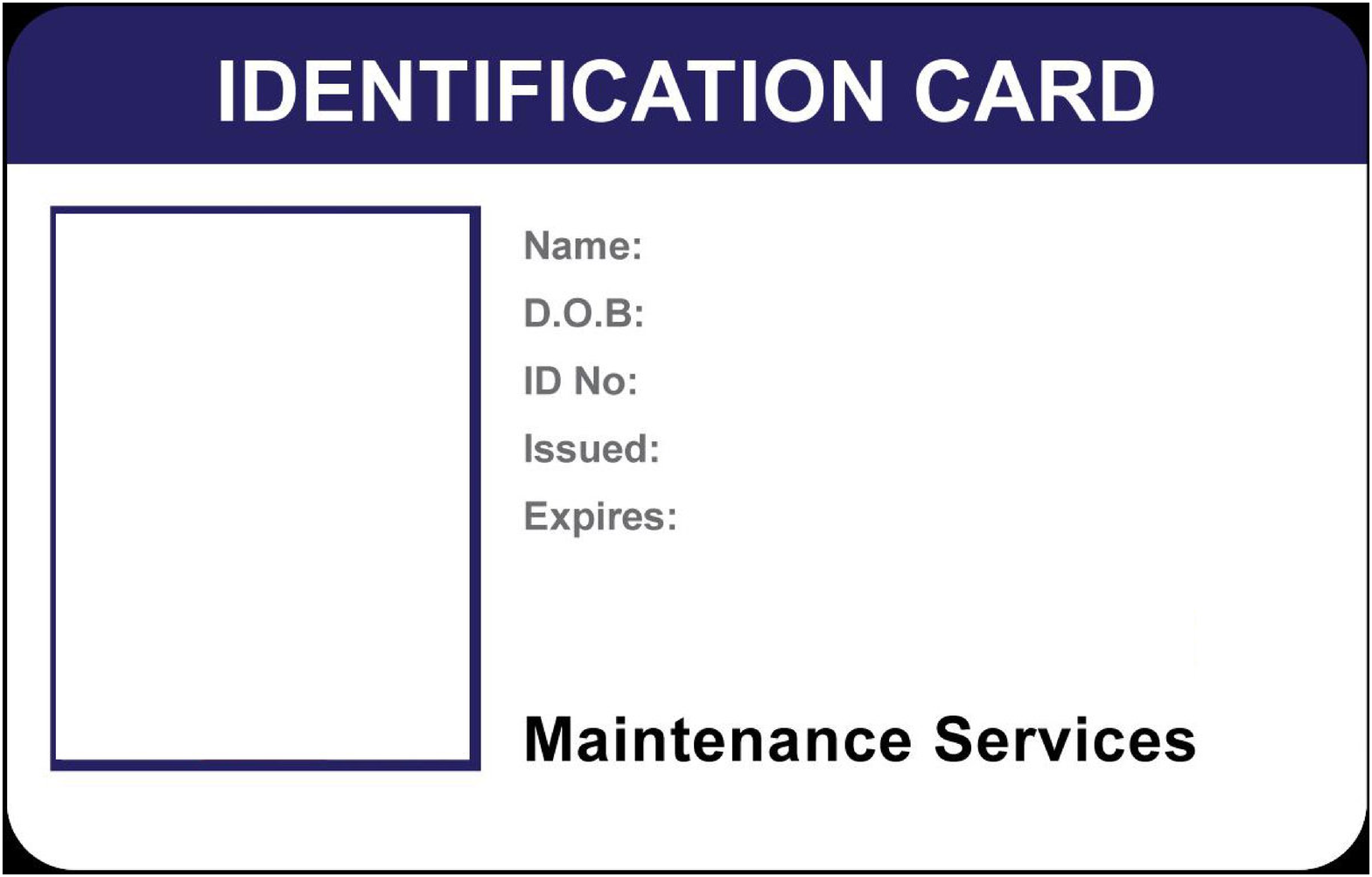}
}
\subfigure[Legal Document]{
\centering
\includegraphics[scale=0.09]{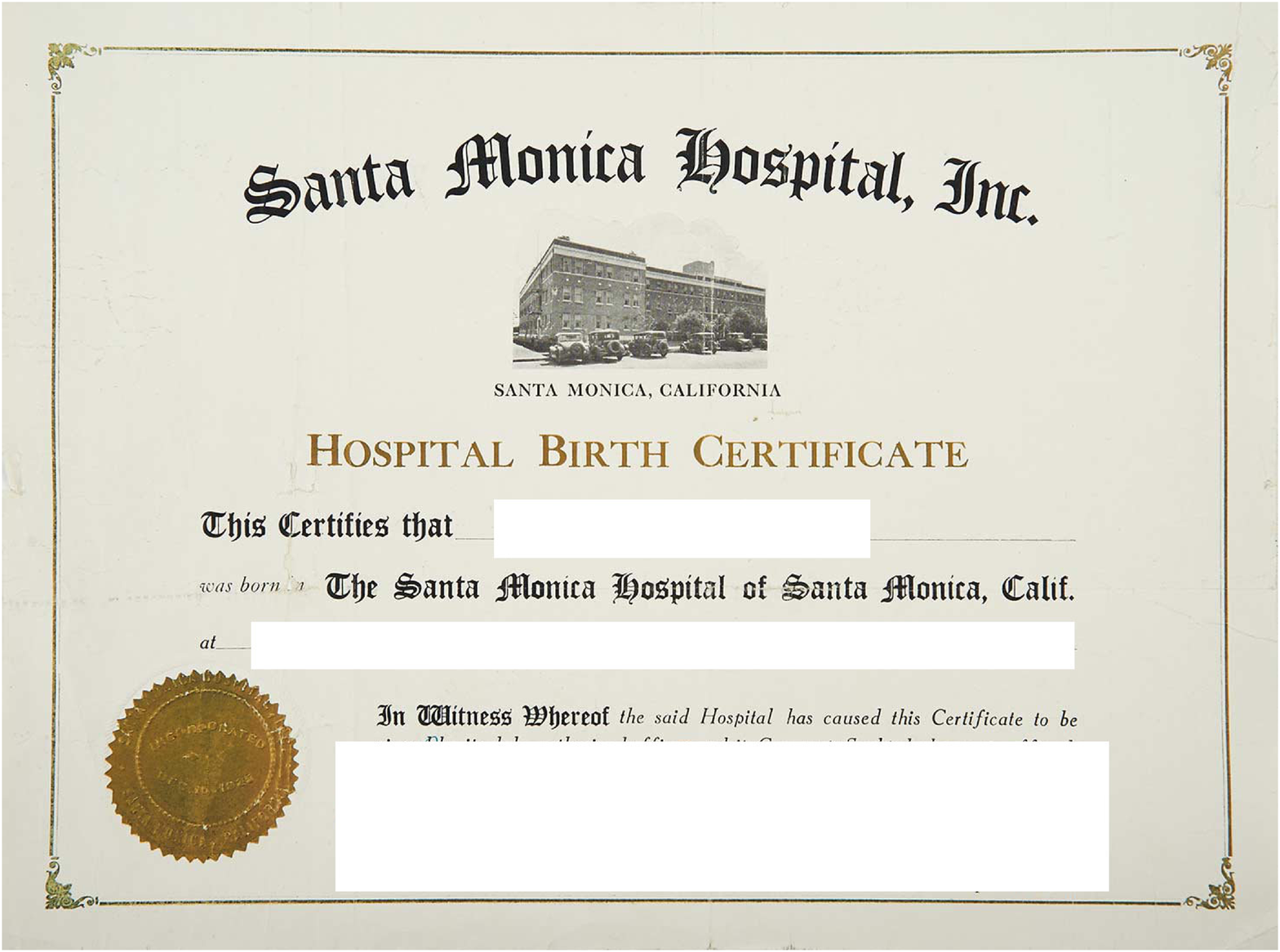}}
\subfigure[Family]{
\centering
\includegraphics[scale=0.12]{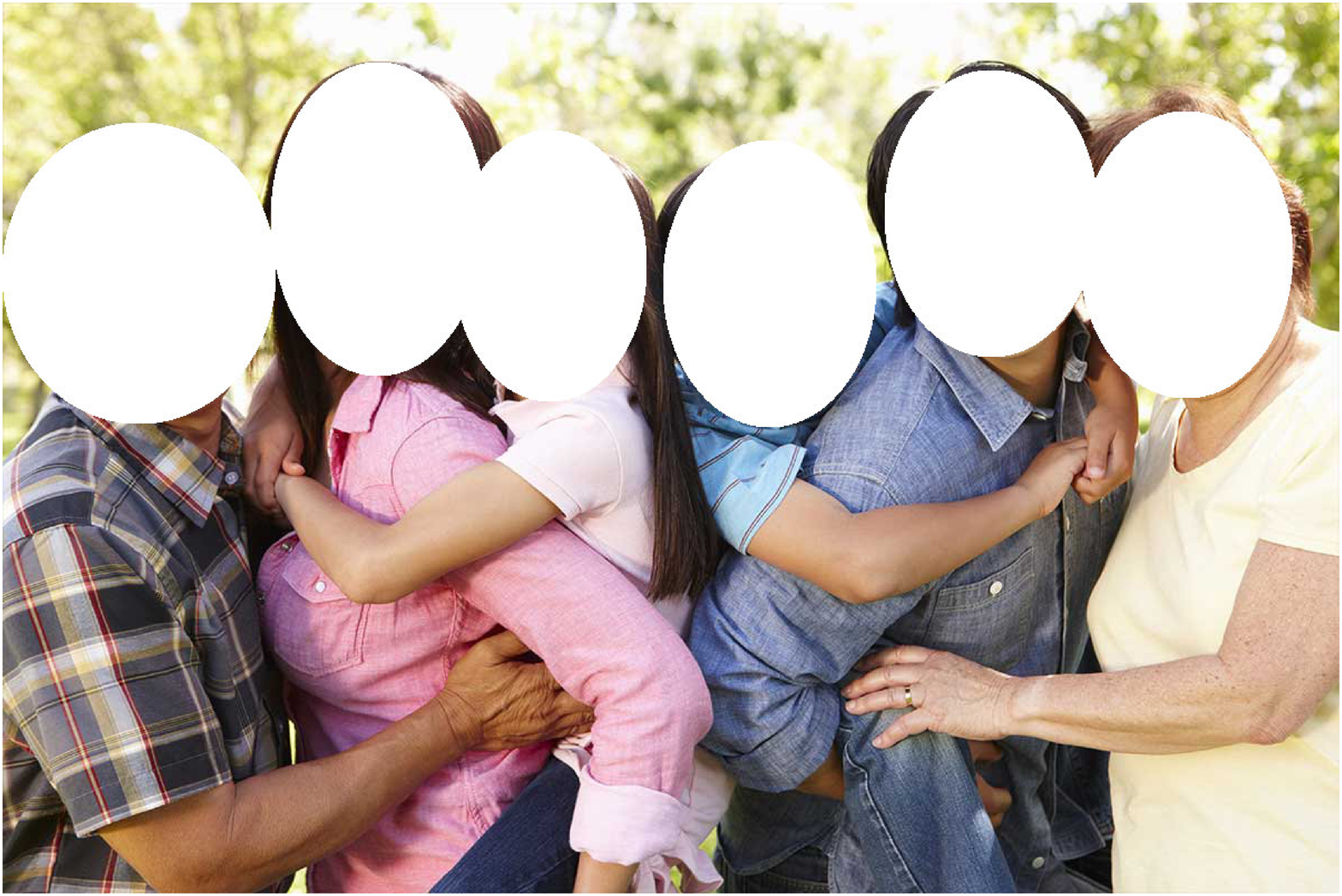}}
\subfigure[Nude]{
\centering
\includegraphics[scale=0.3]{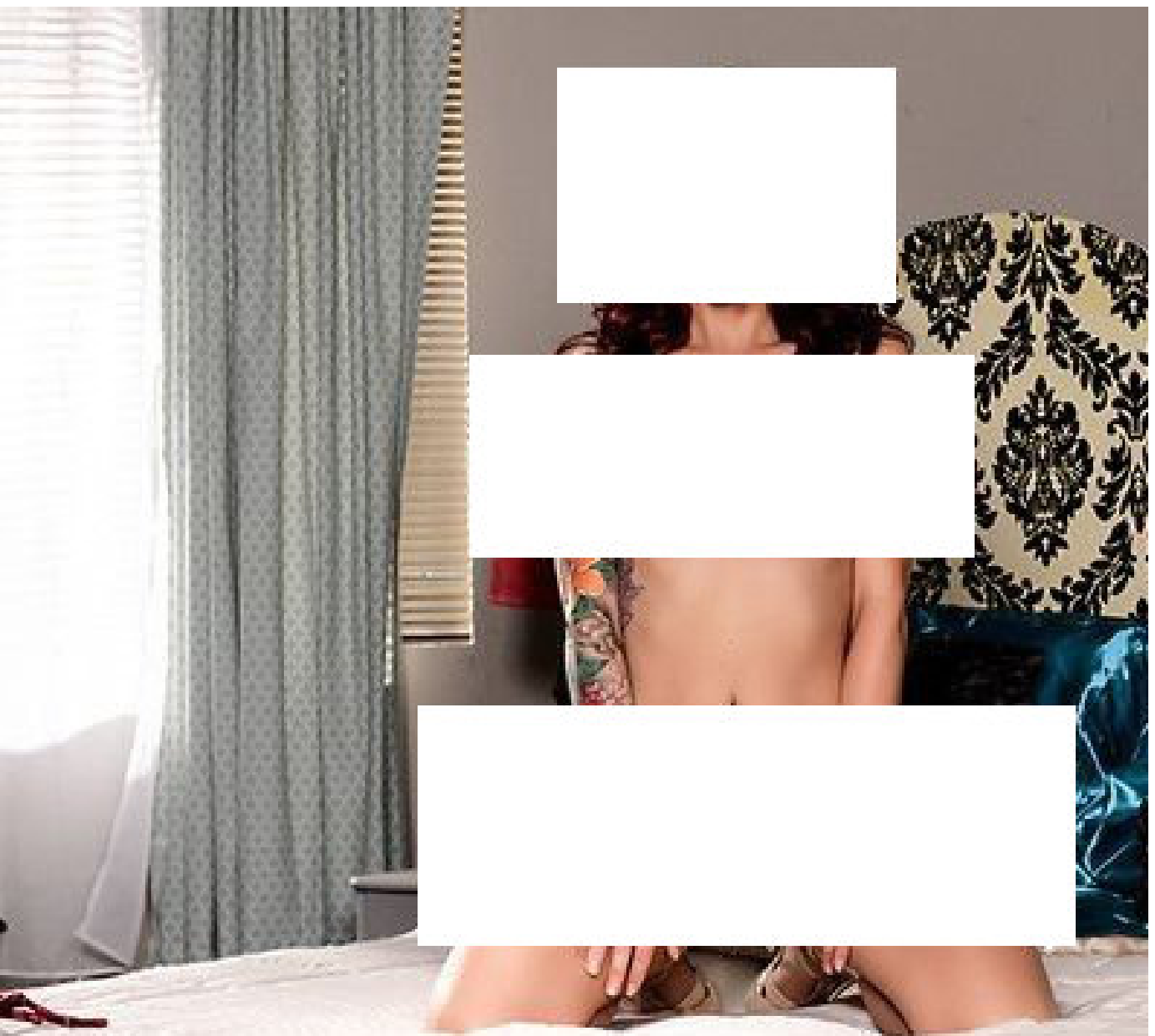}}
\subfigure[Public]{
\centering
\includegraphics[scale=0.45]{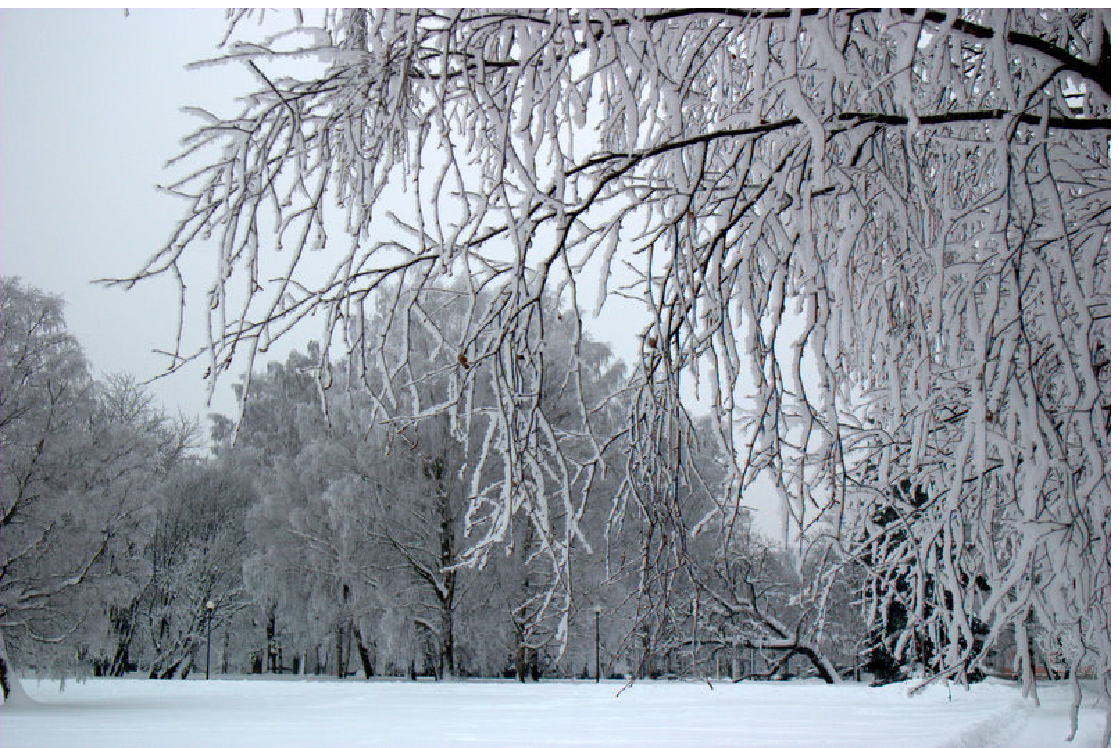}}
\caption{Example photos in our dataset. Sensitive information is removed from the photos.}\label{fig:example}
\end{figure}

Since private photos that are shared on a public domain are limited, it is a challenging task to collect private photos for training deep learning models. Furthermore, there is no standard definition of `private photo' applicable for every user, since it is a very subjective determination. Thus it is hard to collect one private photo dataset to cover all cases. However, Zerr et al. \cite{zerr2012privacy} published a dataset collected from Flickr, which is the only known publicly available dataset for photo privacy research at this time. This dataset consists of 37,535 photos, which are labeled as Private, Public and Undecided. Since the private photos in this dataset do not include most of the private types reported from our survey, we only use the Public photos from this dataset as public photos in our dataset. Additionally, we collect 3,097 private photos in four common types as shown in Table \ref{tb:content} from Google Image, with some example photos shown in Figure \ref{fig:example}. 80\% of the dataset is used for training, and the remained 20\% is used for testing. The distribution of each type of photo is illustrated in Table \ref{tb:distribution}.

\begin{table}[h]
	\centering
	\caption{Dataset Distribution}
	\resizebox{0.49\textwidth}{!}{
	\begin{tabular}{c|c}
		\hline
		\textbf{Photo Type} & \textbf{Number of Photos}\\
		\hline
		Photo ID (e.g., driver's license, passport) & 1353 \\
		Legal Documents (e.g. SSN) & 469 \\
		Family (e.g., family party) & 682\\
		Nudes & 543\\		
		Public & 14664\\
		\hline
	\end{tabular}}\label{tb:distribution}
\end{table}

\subsection{Classification Accuracy}\label{sub:accuracy}
As described above, we trained MobileNets models with a fixed input dimension of $224\times224$ but with different width multipliers. To be specific, we set the width multiplier as 1.0, 0.75 , 0.5 and 0.25 separately. To compare the classification accuracy, we compare our MobileNets model with two baseline models.  One is the Inception\_v3 model \cite{szegedy2016rethinking} that is trained on the same dataset. The other one is that we extract the bag of visual words (BOVW) \cite{yang2007evaluating} from photos and then train it with support vector machine (SVM). All the models except SVM+BOVW are trained in two ways. Firstly, we directly train each model with our dataset. Secondly, we adopt transfer learning to train each model; i.e., we firstly train each model with ImageNet and then fine-tune the model with our dataset. 

As shown in Table \ref{tb:accuracy}, the classification accuracy of each model that is trained in transfer learning is higher than that of each directly trained model. In particular, we observe that the accuracy improves between 17\% and 23\% (in absolute value). It also shows the Inception\_v3 model has a slightly higher accuracy than the MobileNet\_1.0\_224 model, but the model size is much bigger than the MobileNet\_1.0\_224 model. This means it requires much more computation resources for only a little performance improvement, which is not a good fit for resource-constrained mobile phones. In addition, with respect to those MobileNets models, with the decreasing width multiplier the classification accuracy becomes lower and the model size is smaller. Based on the above comparisons, we choose the MobileNet\_1.0\_224 model with transfer learning as our final classifier due to its high classification accuracy and reasonable model size. 

\begin{table}[h]
	\centering
	\caption{Comparison of Classification Accuracy and Model Size}
	\resizebox{0.49\textwidth}{!}{
	\begin{tabular}{c|c|c}
		\hline
		\textbf{Model} & \textbf{Accuracy} & \textbf{Model Size}\\
		\hline
		SVM+BOVW & 73.2\% & 400 MB\\
		Inception\_v3 & 80.3\% & 87.4 MB \\
		MobileNet\_1.0\_224 & 77.5\% & 17.1 MB\\
		MobileNet\_0.75\_224 & 72.6\% & 10.5 MB\\
		MobileNet\_0.5\_224 & 68.7\% & 5.5 MB\\
		MobileNet\_0.25\_224 & 63.3\% & 2 MB\\
		Inception\_v3+Transfer Learning & 97.3\% & 87.5 MB\\
		MobileNet\_1.0\_224+Transfer Learning & 94.3\% & 17.1 MB\\
		MobileNet\_0.75\_224+Transfer Learning & 93.5\% & 10.5 MB\\
		MobileNet\_0.5\_224+Transfer Learning & 91.2\% & 5.5 MB\\
		MobileNet\_0.25\_224+Transfer Learning & 89.7\% & 2 MB\\
		\hline
	\end{tabular}\label{tb:accuracy}}
\end{table}

As shown in Table \ref{tb:topaccuarcy}, we also evaluate the classification accuracy for each type of private photo. It can be seen that our deep learning model achieves higher classification accuracy for each type of private photo than that of the baseline model. For instance, the classification accuracy of `Photo ID' and `Legal Document' is as high as 97.8\%.
\begin{table}[h]
	\caption{Classification Accuracy for Each Type of Photo}
	\centering
	\resizebox{0.49\textwidth}{!}{
		\begin{tabular}{c|c|c|c|c|c}
			\hline
			\multicolumn{1}{c|}{\textbf{Model}} & \textbf{Photo ID} & \multicolumn{1}{c|}{\textbf{Legal Document} }& \textbf{Family} & \textbf{Nude} & \textbf{Public}\\ 
			\hline
			SVM+BOVW & 75.1\% &  71.5\% & 72.7\% & 63.8\% & 61.5\%\\
			MobileNet\_1.0\_224  & 97.8\%& 97.8\% & 95.6\% & 86.2\% & 94.3\% \\
			\hline
	\end{tabular}}\label{tb:topaccuarcy}
\end{table}

In Table \ref{tb:confusionmatrix} we further explore the misclassifications. Even though there are a small number of misclassifications on each type of private photo, none of these are mistakenly classified as `Public'. That means although there exist such misclassifications on some photos, this will not prevent the PhotoSafer from prompting alerts to user.
\begin{table}[h]
	\centering
	\caption{Confusion Matrix of Private Photo Classification}
	\resizebox{0.49\textwidth}{!}{
	\begin{tabular}{cc|c|c|c|c|c|}
	\cline{3-7}
	& & \multicolumn{5}{c|}{\textbf{Prediction}} \\
	\cline{3-7}
	& & Photo ID & Legal Document & Family & Nude & Public\\
	\hline
	\multicolumn{1}{|c|}{\multirow{4}{*}{\rotatebox[origin=r]{90}{\textbf{Actual}}}} & Photo ID & 265 & 4 & 2 & 0 & 0 \\
	\cline{2-7}
	\multicolumn{1}{|c|}{} & Legal Document & 2 & 92 & 0 & 0 & 0\\
	\cline{2-7}
	\multicolumn{1}{|c|}{} & Family & 4 & 0 & 130 & 2 & 0\\
	\cline{2-7}
	\multicolumn{1}{|c|}{} & Nude & 5 & 0 & 10 & 94 & 0 \\
	\hline
\end{tabular}}\label{tb:confusionmatrix}
\end{table}

\subsection{Photo Content Classification Time}
As presented in Section \ref{sec:systemdesign}, in order to avoid affecting user experience, we store the classification results of all stored photos in a photo content database. We measure the time for retrieving one record of a specific photo from the database, compared with the time for making a classification of that photo in real time. We run a total of 10 trials for each of 100 randomly selected photos, and the average time cost is described in Table \ref{tb:time}. It shows the time cost of the database-based approach is 38 time less than that of running real-time classifications.

\begin{table}[h]
	\centering
	\caption{Time Cost for Obtaining Photo Content Classification Result: From Database vs. Real-Time Classification }
	\begin{tabular}{c|c}
		\hline
		\textbf{Method} & \textbf{Time} \\
		\hline
		From Database & 5.2 ms \\
		Real-Time Classification & 190.7 ms \\
		\hline
	\end{tabular}\label{tb:time}
\end{table}

\section{Discussions}\label{sec:discussion}
Even though we implemented a prototype system, we did not integrate it into the Android kernel due to time limitation. In addition, although the proposed system is designed for Android platform, it may also be applicable to other mobile platforms. This section discusses the limitations of the current design and implementation, and how it can be improved in the next step of research.

\textbf{Kernel Interposition.}\hspace{0.1in} In our current prototype system, we implemented a function to simulate apps that may access photos under different system status and app-running status in the real world. This is the only way the prototype can interpose photo access and monitor which specific photo is being accessed. Otherwise, the prototype system requires the root privilege in the Android system, which is not safe for user to install such software. However, as described in Section \ref{sec:systemdesign}, the best way is to implement photo access interposition in the Android kernel. Since photos are accessed as regular files in Android, all file access should be interposed. Additionally, the kernel interposition needs to determine if the accessed file is a photo through checking file extension (e.g., .jpg), so that the system can decide in the kernel whether the accessed file needs our proposed access control.

\textbf{App Whitelist.}\hspace{0.1in} In the current system design, we determine whether a photo access by a specific app is `unauthorized' based the system status and app-running status, which can cover most cases. However, in some cases, users are satisfied with some apps that are running in the background access photos. For instance, some users allow the Google Photos app to backup the stored photos even if it is running in the background. To consider such cases, the current system design can be improved by adding an app whitelist. It allows users to specify which apps can be granted access to all stored photos without the proposed access control.

\section{Related Work}\label{sec: relatedwork}
\textbf{Android Permissions}\hspace{0.1in} The Android platform offers users two approaches to controlling permissions. Before Android 6.0 (Marshmallow), apps are required to disclose the full list of resources that they want to access at installation. Users must grant all requested permissions; otherwise the installation will be discontinued. Some work \cite{felt2012android,kelley2012conundrum} has shown that few users pay attention to and really understand the meaning of installation-time permissions. After Android 6.0, users need to grant permissions only when an app requests a sensitive resource for the first time. This scheme can offer users contextual clues about why the requested resource is necessary for an app. However, it does not account for the fact that the user's preference for subsequent permission requests might be changed under different contextual circumstances. 

Work has been done on permission models \cite{felt2012android,felt2012ve,egelman2013choice,sadeh2014modeling,shklovski2014leakiness}, which found that users usually do not know how apps access sensitive resources and how such access is managed. Shih et al. \cite{shih2015privacy} showed that private information is more likely to be leaked when users are unaware of the purpose for requesting a specific sensitive resource.

Almuhimedi et al. \cite{almuhimedi2015your} analyzed AppOps, which is a permission manager introduced in Android 4.3 but removed in Android 4.4.2. AppOps allows users to review and modify app permissions after apps are installed. They provided both qualitative and quantitative evidence that the permission manager can increase users' awareness of privacy risks. A new permission management system was introduced in  Android 6.0 to replace AppOps, which allows users to review all permissions that an app has been granted. However, since it is hidden in the deep level of the \texttt{Settings} app,  it is not easy for average users to discover it. There exist several third-party permission management apps, such as XPrivacy \cite{bokhorst2013xprivacy}, DonkeyGuard \cite{donkeyguard}, Permission Manager \cite{permissionmaster} and Privacy Guard \cite{lineageos}. However, these apps require additional privileges to support their functionalities, since there is no official approach offered to third-party apps to modify the permission system. For example, XPrivacy requires an unlocked bootloader and a custom recovery partition. Such restrictions are needed to protect the permission system against interfering by malicious apps.

\textbf{Photo Privacy}\hspace{0.1in} Ra et al. \cite{ra2013p3} designed a system P3 to protect privacy of photos when they are shared on online social networks. He et al. \cite{he2016puppies} proposed an approach to protect users' privacy for photo sharing. Jana et al. \cite{jana2013scanner} proposed a system Darkly based on the OpenCV library, which protects users' private information from continuous-sensing applications. Templeman et al. \cite{templeman2014placeavoider} implemented a system PlaceAvoider to protect visual privacy by identifying sensitive places in video streams. Li et al. \cite{li2016privacycamera,li2018politecamera} proposed two systems for protecting bystanders' privacy in photo taking. Tan et al. \cite{tan2014short} designed an access control scheme to protect private photos on mobile phones, but it depends on pre-specified target faces on mobile phones and can only provide limited protection. Zerr et al. \cite{zerr2012privacy} collected a photo dataset from Flickr with labels public, private or undecided. Then, they extracted low-level features and trained a SVM model to identify private photos. Squicciarini et al. \cite{squicciarini2014analyzing} conducted an extensive study based on the Flickr dataset collected by Zerr et al. and developed learning models to estimate adequate privacy settings for shared photos in online social networks. Similar to the work by Zerr et al. \cite{zerr2012privacy}, Liu et al. \cite{liu2011analyzing} recruited workers to label photos collected from Facebook as shared with ``only me'', ``some friends'', ``all friends'', ``friends of friends’’, and ``everyone''. They found that there is a big difference between the actual labels on Facebook and labels obtained from workers. The difference is due to the reason that Facebook users usually share photos using the default privacy setting. Such difference indicates that the default privacy setting on Facebook is much lower than the privacy protection that users desire.

\section{Conclusion and Future Work}\label{conclusion}
Motivated by a user survey and analysis of 200 apps' permission requests, both of which showed the potential risk of private photos being leaked without being known, we proposed a system PhotoSafer to protect private photos that are stored on mobile phones from being accessed without users' awareness. The access control on those private photos is enforced by checking the system status and photo content. A mobile-compatible private photo classifier was designed with transfer learning. We implemented a prototype system, and evaluated its performance and cost through experiments.

In future work, we plan to implement the app whitelist module and integrate the system into the Android kernel. The current design can protect several common types of private photos. However, photo privacy is a very subjective problem. Thus we will also consider designing personalized systems to protect user-dependent private photos.

\bibliographystyle{IEEEtran}
\bibliography{reference}
\end{document}